\documentclass[10pt, letterpaper]{article}

\usepackage[OT1]{fontenc}
\usepackage[in]{fullpage}
\usepackage{titlesec}

\usepackage[pdftex,pdfpagemode={UseOutlines},bookmarks,bookmarksopen,colorlinks,linkcolor={black},citecolor={black},urlcolor={red}]{hyperref}
\usepackage{graphicx}
\usepackage{bm}
\usepackage{amsmath,amssymb}
\usepackage{aas_macros}

\graphicspath{{figure/}}

\usepackage{mathptmx}  

\titleformat*{\section}{\normalsize\bfseries\scshape}
\titleformat*{\subsection}{\normalsize\bfseries\scshape}
\titleformat*{\subsubsection}{\normalsize\bfseries\scshape}
\titleformat*{\paragraph}{\normalsize\bfseries\scshape}
\titleformat*{\subparagraph}{\normalsize\bfseries\scshape}

\newcommand{\half}{\frac{1}{2}}

\newcommand{\xv}{\mathbf{x}}

\newcommand{\thetav}{\boldsymbol\theta}
\newcommand{\rhov}{\boldsymbol\rho}

\newcommand{\data}{\mathbf{d}}
\newcommand{\ntel}{n_{\rm telescopes}}

\newcommand{\orbitparams}{\thetav}
\newcommand{\telpos}{\xv}
\newcommand{\telorient}{\boldsymbol \gamma}
\newcommand{\noisevari}{\sigma^{2}_{{\rm noise},i}}
\newcommand{\model}{\mathbf{m}}
\newcommand{\fc}{\mathbf{H}}
\newcommand{\fcx}{H_x}
\newcommand{\fcy}{H_y}
\newcommand{\psf}{\Pi}
\newcommand{\psfwidth}{\sigma_{\psf}^{2}}
\newcommand{\streak}{\boldsymbol \ell}
\newcommand{\streakim}{\hat{\streak}}
\newcommand{\rect}{{\rm rect}}
\newcommand{\sinc}{\rm sinc}
\newcommand{\erf}{\rm Erf}
\newcommand{\modelparams}{\omega}

\newcommand{\nepochs}{N_{\rm epochs}}

\newcommand{\pr}{{\rm Pr}}
\newcommand{\normdist}{\mathcal{N}}

\begin{document}
  
\title{\large{{\bf Synthesis of disparate optical imaging data for space 
domain awareness}}}

\author{
\large{{\bf Michael D. Schneider \& William A. Dawson}}\\
\large{{\it Lawrence Livermore National Laboratory, 
P.O. Box 808 L-211, Livermore, CA 94551-0808, USA.}}\\
\vspace{0.01in}\\
\normalsize{LLNL-CONF-836209}
}

\date{}

\maketitle

\section{Abstract}
We present a Bayesian algorithm to combine optical imaging of unresolved objects
from distinct epochs and observation platforms for orbit determination and 
tracking. By propagating the non-Gaussian uncertainties we are able to optimally combine imaging 
of arbitrary signal-to-noise ratios, allowing the integration of data from 
low-cost sensors. Our Bayesian approach to image characterization also allows 
large compression of imaging data without loss of statistical information. With
a computationally efficient algorithm to combine multiple observation epochs
and multiple telescopes, 
we show statistically optimal orbit inferences.

\section{Introduction}
\label{sec:introduction}

The Geosynchronous Earth Orbit (GEO) is an increasingly crowded environment
requiring regular monitoring for orbit determinations and refinements in 
catalog maintenance. When using optical telescopes to monitor GEO, many 
observations are needed to measure sufficient arc length to constrain an orbit.
The exposure time for a single exposure is often limited by the requirement to 
avoid detector saturation from bright stars. Or, in the case of non-sidereal tracking, the exposure time is limited by the avoidance of star streaks covering too much detector area. The telescope aperture then limits
the faintest magnitudes that can be tracked, which are correlated with satellite 
or debris size~\cite{schildknecht2004}.

We present an algorithm for characterization of streaks in optical CCD imaging 
of arbitrary signal-to-noise ratio (assuming an initial detection has been made
\cite{dawson_amos2016})
and subsequent orbit determinations including uncertainty propagation and 
combinations of information from multiple observing epochs or telescopes. We 
characterize streak information (position, length, and flux) via a statistical 
model that allows statistical compression of image information.

\section{Method}
\label{sec:method}

\subsection{Image forward model}

We seek an analytic model for a streak image
because with parameters that we can infer in a probabilistic framework. This in 
principle allows us to optimally utilize all the information in the streak image, including the positions, flux, and point-spread function (PSF) size.  
An analytic model including the PSF also avoids computing time that might be spent in numerical convolutions of the image with the PSF when rendering an image model.

We choose a Gaussian model for the PSF because the Fourier transform is also a
Gaussian and the central peak of an Airy disk (the diffraction limited PSF of an
unaberrated system with a circular aperture) is well approximated by a Gaussian.
\begin{equation}
    \psf_i(\fc_i) \equiv \exp\left(-\half |\fc_i|^{2} / \psfwidth\right),
\end{equation}
where $\fc_i$ are angular positions in the focal plane of the $i$th telescope.

We model a streak as a product of a narrow Gaussian (representing the width that can be
made arbitrarily small) and a rectangular window (representing the extent of the streak).
We further assume that the streak model is evaluated in a coordinate system with the center
of the streak at the origin and the extent of the streak along the $x$-axis. This can always
be accomplished by means of a coordinate translation by $x_0,y_0$ and a rotation by $\phi_0$ giving observed focal plane coordinates $\mathbf{H}$,
\begin{align}
    \fc_x &= \cos\phi_0 (x - x_0) - \sin\phi_0 (y - y_0)
    \\
    \fc_y &= \sin\phi_0 (x - x_0) + \cos\phi_0 (y - y_0).
\end{align}
Then (see also \cite{1538-3873-124-921-1197}),
\begin{equation}
    \streak(\fc) = \ell_0
    \rect\left(\frac{\fcx}{L}\right)
    \exp\left(-\half \frac{\fcy^2}{\sigma_y^2}\right),
\end{equation}
where $L$ is the length of the streak, $\ell_0$ is the surface brightness of 
the streak (i.e., luminosity per unit area),
and $\rect$ is the rectangular window function.

The Fourier transforms of the PSF and streak model are,
\begin{equation}
    \tilde{\psf}_{i}(\rhov) = 
    \exp \left(-\half|\rhov|^{2} \psfwidth\right),
\end{equation}
and,
\begin{equation}
    \tilde{\streak}(\rhov) =
    \ell_0\, \sinc \left(\frac{L \rho_x}{2}\right)
    \exp\left(-\half \rho_y^2\sigma_y^2\right).
\end{equation}

The streak image after convolution with the PSF is,
\begin{equation}\label{eq:streakimage}
    \streakim(\fc) =
    \frac{\ell_0}{2}
    \left[
    \erf\left(\frac{L-2\fcx}{2\sqrt{2\psfwidth}}\right)
    +
    \erf\left(\frac{L+2\fcx}{2\sqrt{2\psfwidth}}\right)
    \right]
    \exp\left(-\half \frac{\fcy^2}{\psfwidth + \sigma_y^2}\right)
    \frac{\sqrt{2\pi\sigma_y^2}}{\sqrt{2\pi(\psfwidth + \sigma_y^2)}}
\end{equation}

\autoref{eq:streakimage} is our model for the image of a streak,
which has parameters,
\begin{equation}
    \modelparams \equiv \left[x_0, y_0, \phi_0, \ell_0, L, \psfwidth\right].
\end{equation}
We will always set $\sigma_y^2 \ll \psfwidth$ for this model.

We fit the streak image parameters $\modelparams$ to the pixel values of a 
CCD image $\data$ by means of a likelihood function, with 
the likelihood for all available images of a unique source assumed to factor 
into a product of the likelihoods for each image.
Assuming also Gaussian, uncorrelated pixel noise in each image,
\begin{equation}\label{eq:image_like}
    \pr(\data | \modelparams_i, \noisevari) =
    \prod_{i=1}^{\ntel}
    \normdist_{\data_i}\left(\model(\modelparams_i), \noisevari\right),
\end{equation}
where $\model$ is the model prediction for the pixel data in telescope $i$ given
streak image parameters $\modelparams_i$,
$\noisevari$ is the per-pixel noise variance in telescope $i$, and $\normdist_x$
indicates a Normal (or Gaussian) distribution for the variable $x$.

Given a single streak image, we infer `interim' posterior constraints on the 
parameters of the streak image using the likelihood function in 
\autoref{eq:image_like},
\begin{equation}
    \pr(\modelparams_i|\data_i, \noisevari) \propto
    \pr(\data_i | \modelparams_i, \noisevari) \pr(\modelparams_i | I_0),
\end{equation}
where $I_0$ encodes the prior assumptions for fitting of single-epoch streak 
images. Sampling in the streak image parameters $\modelparams_i$ provides a 
convenient way to extract source information and compress the image data into a
set of $K$ model parameter samples $\modelparams_{i;k}, k=1,\dots,K$.

\subsection{Preliminary orbit determination}

To obtain preliminary orbit determinations (PODs), we want to sample from 
$\pr(\theta|\data)$, where $\theta$ are the orbit parameters.

The streak image parameters are
\begin{equation}
    \omega = (x_0,y_0,L,\phi_0,\ell_0,\sigma^2_{\Pi})
    \equiv (\hat{\omega}, \ell_0,\sigma^2_{\Pi}),
\end{equation}
where we have split out $\ell_0,\sigma^2_{\Pi}$ as nuisance parameters for
orbit determination.

The (osculating) Keplerian elements $\theta$ are entirely determined given the 
remaining streak image parameters $\hat{\omega}$ coupled with ranges, $\rho$, 
(i.e., line-of-sight distances) at the exposure start and end 
times~\cite{shefer2010},
\begin{equation}\label{eq:var_transform}
    \theta = \theta(\hat{\omega}, \rho, t).
\end{equation}
Note there is an ambiguity in the variable transformation in 
\autoref{eq:var_transform} if the orbit is not known to be prograde or 
retrograde. In this case, both solutions must be tried. This degeneracy can
be broken if subsequent exposures are taken or a time-tagging procedure as in 
\cite{Park:2013hr} is used.

The posterior for the POD is then,
\begin{equation}\label{eq:pod_posterior}
    \pr(\theta|\data, \alpha, t)\,d\theta \propto
    \pr(\theta(\rho,\hat{\omega},t))
    \int d\ell_0 d\sigma^2_{\Pi}\, 
    \left[\pr(\data|\omega, t) \pr(\omega|t)\right]
    \pr(\rho|t,\alpha)
    \frac{1}{\pr(\omega|t)}
    \left|\frac{d(\hat{\omega}, \rho)}{d\theta}\right|\, d\theta,
\end{equation}
where the bracketed expression $\pr(\data|\omega, t) \pr(\omega|t)$ is the posterior from which samples
of $\omega$ are drawn in the interim sampling to fit the streak image model,
$\pr(\rho|t,\alpha)$ is an asserted prior distribution for the ranges, 
and $\alpha$ represents the parameters describing the prior distribution. We 
must assert the range prior because observations of a single streak cannot 
constrain the range parameters.

\autoref{eq:pod_posterior} suggests a sampling algorithm for POD,
\begin{enumerate}
    \item Select $N$ (uncorrelated) samples of $\hat{\omega}_i$ from the interim
    sampling used to fit the streak image model,
    \item Draw $N$ samples of $\rho_i$ from $\pr(\rho|t,\alpha)$ given
    hyper-parameters $\alpha$ (as in \cite{schneider12}),
    \item Compute $\theta_i(\rho_i, \hat{\omega}_i, t)$ for each $i=1,\dots,N$,
    \item Optionally, re-weight or compute accept/reject sampling of $\theta_i$
    under a new prior,
    \begin{equation}\label{eq:pod_prior}
        p \equiv \pr(\theta) \left|\frac{d(\hat{\omega},\rho)}{d\theta}\right|
        \frac{1}{\pr(\hat{\omega}|t)}.
    \end{equation}
\end{enumerate}

\subsection{Combining preliminary orbits}
After inferring PODs for each observation epoch separately, we have a set of 
posterior samples, $\orbitparams^{(i)}$, for each epoch or telescope $i$. To 
refine our orbit inferences, we describe a method to re-weight the samples 
$\orbitparams^{(i)}$ to propagate the information from the combination
of all observations.

We use the Generalized Multiple Importance Sampling (MIS) algorithm from \cite{mis}.
The MIS weights associated with the $k$th interim posterior sample 
$\theta^{(i)}_{k}$ are,
\begin{equation}\label{eq:mis_weights}
    w_{k}^{(i)} \equiv \frac{\prod_{j=1}^{\nepochs} \pr(\data_{j}|\theta^{(i)}_{k})}
        {\sum_{j=1}^{\nepochs} \pr(\data_{j}|\theta^{(i)}_{k})},
\end{equation}
where the likelihood of the single-epoch data $\data_j$ is implicitly defined 
by the integrand in \autoref{eq:pod_posterior}. The key feature to note in 
\autoref{eq:mis_weights} is that we evaluate the likelihood of the orbit 
parameters from the epoch $i$ analysis given the data from all other epochs $j$.
The weights in \autoref{eq:mis_weights} are thus only large if the likelihoods 
of all available images are consistent with the given parameters 
$\orbitparams^{(i)}$.

To evaluate the single-epoch likelihoods in \autoref{eq:mis_weights},
\begin{enumerate}
    \item Convert $\orbitparams^{(i)}_{k}$ to 
    $\hat{\modelparams}^{(i)}_{k}(\orbitparams, \telpos_j, \telorient_j)$
    and $\rho^{(i)}(t_j)$ (where $\telpos_j$ represents the telescope position
    and $\telorient_j$ represents the pointing vector),
    \item Sample $\ell_0, \psfwidth$ 
    (and, optionally, $\telpos_j$ and $\telorient_j$)
    with replacement from the interim image model samples from epoch $j$,
    \item Evaluate
    \begin{equation}
        \pr(\data_j|\omega^{(i)}, t_j) \pr(\rho^{(i)}|t_j,\alpha)
        p_{j},
    \end{equation}
    where $p_{j}$ is defined in \autoref{eq:pod_prior}.
\end{enumerate}
We have also found that a kernel density approximation of 
$\pr(\data_{j}|\theta^{(i)})$ can be sufficient for accurate evaluation of 
\autoref{eq:mis_weights}.

\section{Results}
\label{sec:results}

\begin{figure}
    \centerline{
        \includegraphics[width=0.5\textwidth]{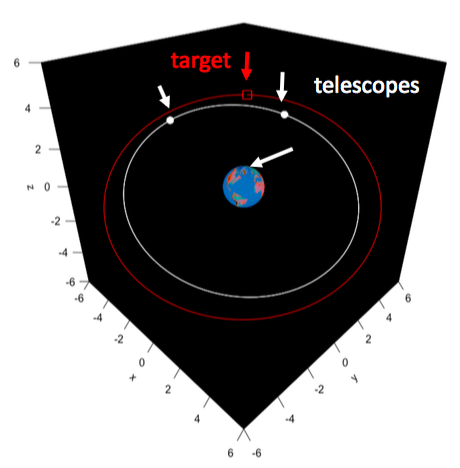}
    }
    \caption{The scenario for our simulation study includes 3 telescopes 
    observing a single GEO satellite. One telescope is ground-based and two 
    are space-based.}
    \label{fig:sim_setup}
\end{figure}

\begin{figure}
    \centerline{
        \includegraphics[width=0.3\textwidth]{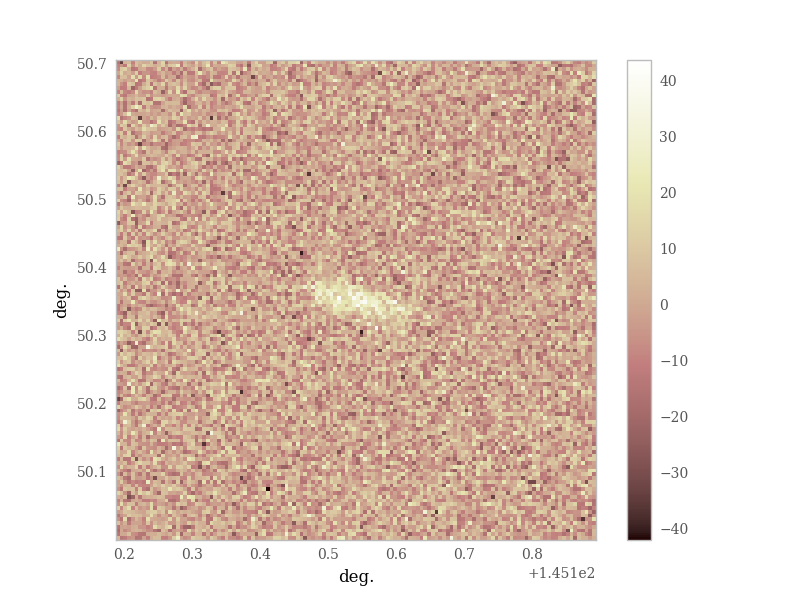}
        \includegraphics[width=0.3\textwidth]{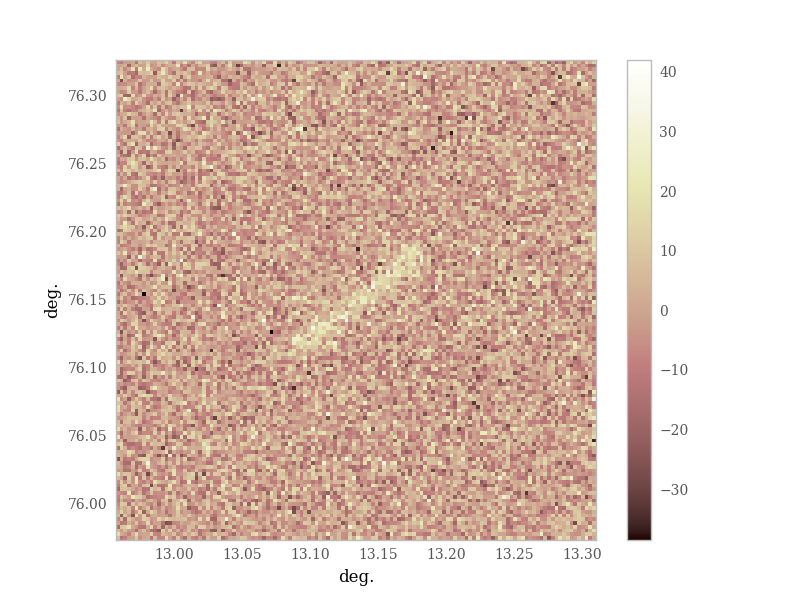}
        \includegraphics[width=0.3\textwidth]{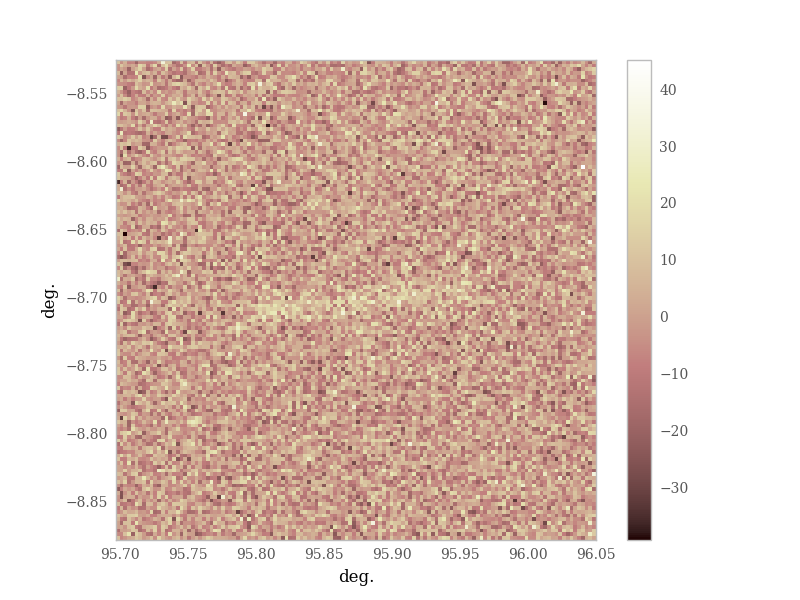}        
    }
    \caption{Mock images of a GEO satellite in a 30 second exposure taken from
    ground (left) and two space-based platforms (middle and right).}
    \label{fig:mock_data}
\end{figure}

In this section we demonstrate the algorithm in \autoref{sec:method} with 
simulated images of a GEO satellite where we can compare with the known true
orbit. We consider a scenario with 3 observatories at different locations: 
1 telescope located on the surface of the Earth and 2 satellites in a sub-GEO 
orbit that are offset in orbital phase by a fraction of a period as in 
\autoref{fig:sim_setup} (see 
\cite{danescu2012long} for a related scenario). 
We will 
compare the orbit inferences from this combination of 3 telescopes to that 
achievable with multiple exposures taken from the single ground-based 
observatory. In all cases we assume 30 second exposures and pixel scales 
of 20 arcseconds on the ground and 10 arcseconds in space. We show the mock 
images in \autoref{fig:mock_data}.

\begin{figure}[!htb]
    \centerline{
        \includegraphics[width=0.8\textwidth]{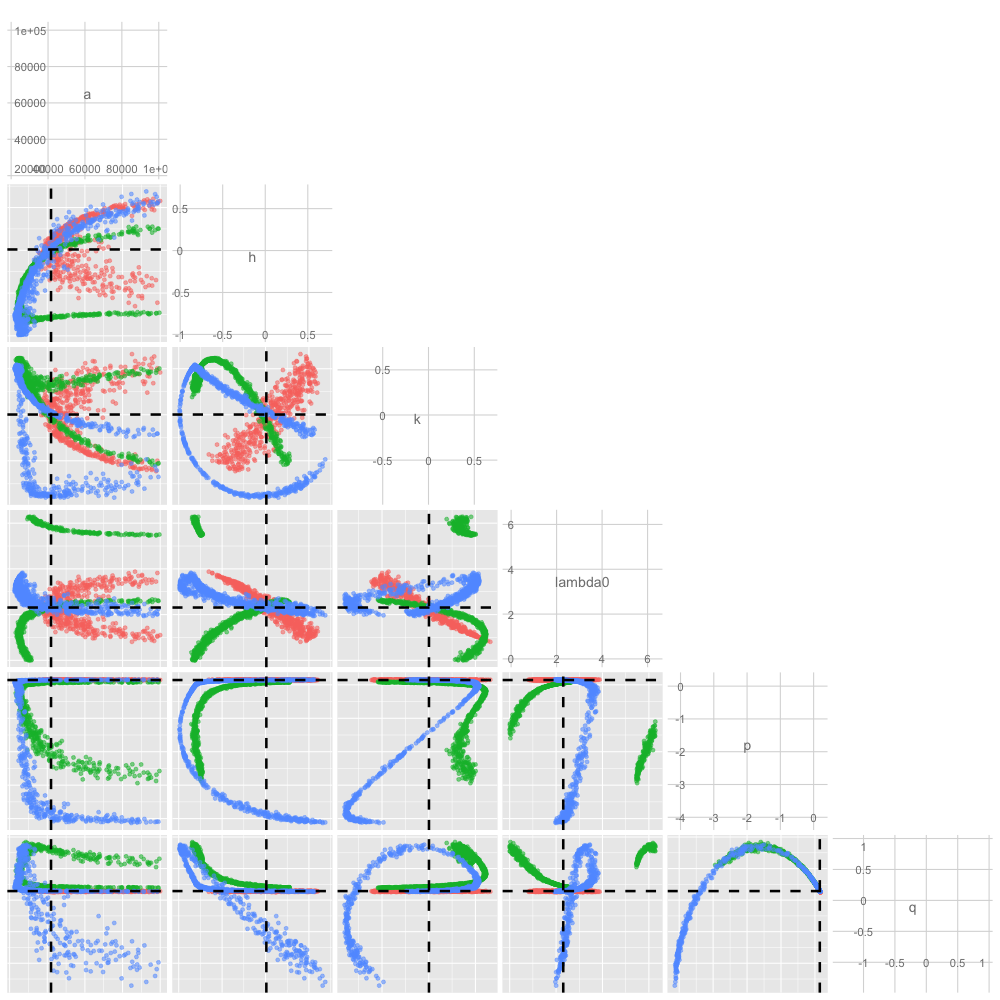}
    }
    \caption{Posterior samples of equinoctial Keplerian orbital elements given 
    a single 30-second observation of a GEO satellite as seen from telescopes 
    either on the ground or in a sub-GEO orbit. 
    Red: ground-based telescope. Blue/green: space-based telescopes offset in 
    orbital phase (see text).}
    \label{fig:thresher_corner}
\end{figure}

In \autoref{fig:thresher_corner} we show the PODs from simultaneous observations
from the ground and space telescopes as 2D marginal samples of the six 
equinoctial Keplerian elements~\cite{broucke72}. Each telescope has a view of 
the satellite from a different location and perspective, which leads to 
vastly different posterior constraints on the equinoctial elements. But, while 
the constraints on the equinoctial elements in \autoref{fig:thresher_corner}
are weak and non-Gaussian for any single telescope, the regions of parameter 
space where the constraints from all telescopes overlap is much smaller. The 
geometric information from the different perspectives on a single satellite thus
helps constrain the orbit.

\begin{figure}[!htb]
    \centerline{
        \includegraphics[width=0.75\textwidth]{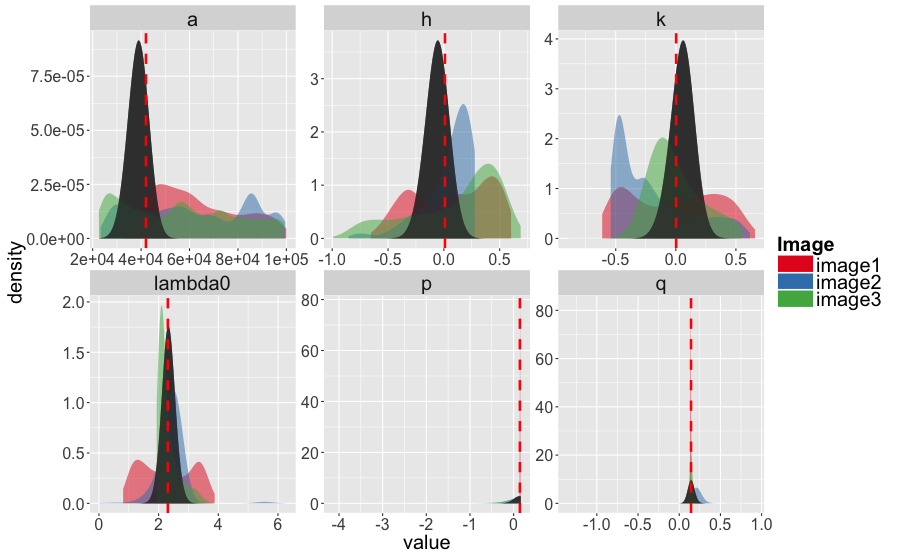}
    }
    \caption{The 1D marginal posterior constraints on the equinoctial Keplerian
    orbital elements from an observing scenario with 1 ground- and 2 
    space-based telescopes taking a simultaneous 30 second observation of a 
    GEO satellite (see text). The red, blue, and green shaded distributions 
    show the posteriors given single-telescope observations (ground or space), 
    while the gray shaded posterior is given the combined information from 
    all 3 telescopes.}
    \label{fig:thresher_marg_dens}
\end{figure}

We show the 1D marginal posterior constraints on the equinoctial elements from 
each telescope individually and combined in \autoref{fig:thresher_marg_dens}.
The combined constraints shown in shaded gray use \autoref{eq:mis_weights} 
to re-weight the posterior samples of the orbital elements from each telescope.
The shaded red, blue, and green distributions in 
\autoref{fig:thresher_marg_dens} show the constraints from each of the 
telescopes individually. 

\begin{figure}[!htb]
    \centerline{
        \includegraphics[width=0.75\textwidth]{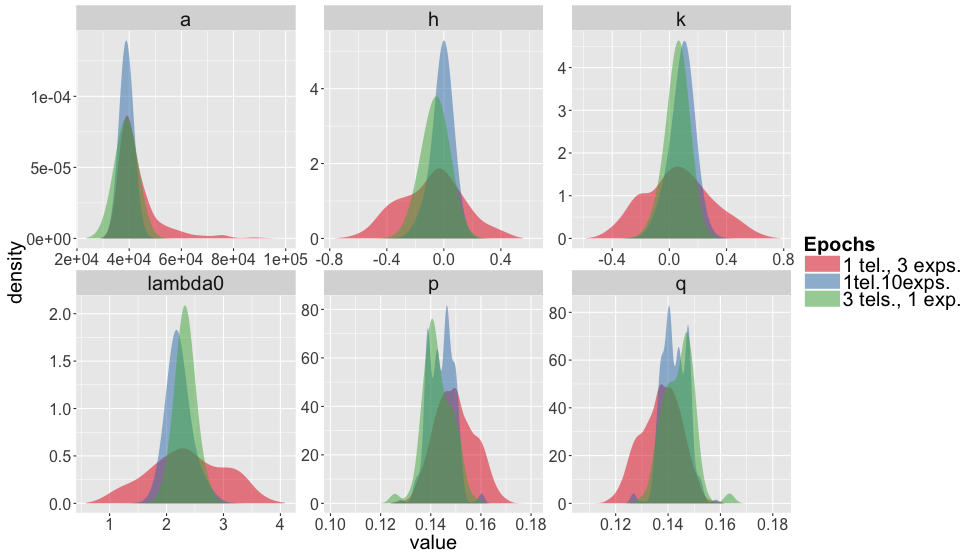}
    }
    \caption{Comparison of the marginal posterior constraints on the 
    equinoctial elements of a GEO satellite from two observing scenarios. 
    Green: constraints from 
    3 telescopes, 1 ground-based and 2 space-based taking a simultaneous 
    30-second observation. Red: 1 ground-based telescope taking 3 consecutive 
    30-second exposures. Blue: same as red, but taking 10 consecutive exposures.}
    \label{fig:thresher_marg_dens_comparison}
\end{figure}

To better understand the impact of the geometric diversity of the combined 
ground- and space-based observatories, we also simulate a series of 
consecutive 30-second exposures from the ground, separated by 5 seconds of 
readout time. In \autoref{fig:thresher_marg_dens_comparison} we compare the 
1D marginal posterior constraints on the equinoctial elements between 
the consecutive ground exposures and the combined ground and space observations
(using only a single simultaneous 30-second exposure on each telescope for the 
latter). 

We see in \autoref{fig:thresher_marg_dens_comparison} that 3 consecutive 
30-second exposures from the ground yields much more uncertain orbit inferences 
than the combination of 30-second exposures from the ground and space. 
However, 10 consecutive exposures from the ground can provide comparable 
orbit inferences to those from the simultaneous 30 second ground and space 
observations.

\begin{figure}[!htb]
    \centerline{
        \includegraphics[width=0.5\textwidth]{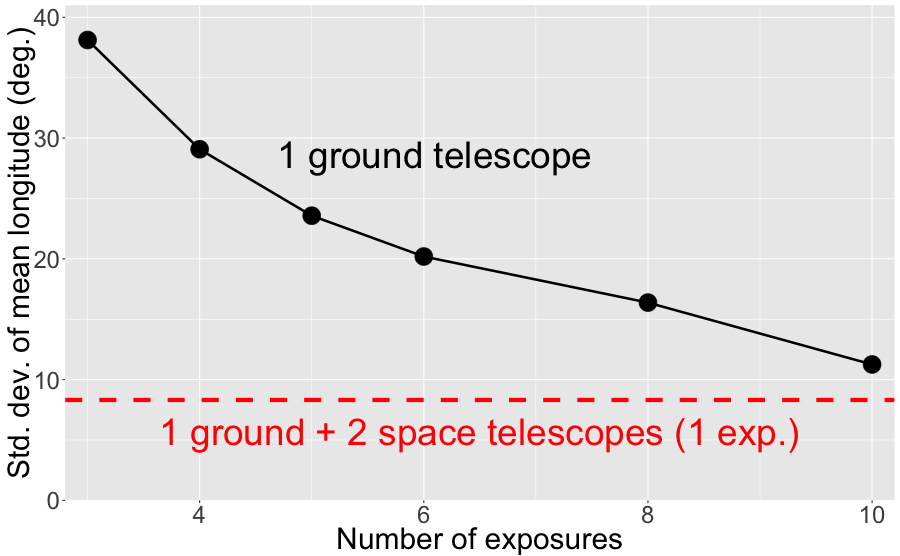}
    }
    \caption{The marginal posterior standard deviation of the mean longitude
    for two observing scenarios. Red dashed horizontal line: a single 
    simultaneous 30-second observation taken from 1 ground- and 2 space-based 
    telescopes (see text). Black points: cumulative constraints from 
    consecutive 30-second exposures (separated by 5-seconds readout) for a 
    single ground-based telescope.}
    \label{fig:thresher_in_track_stddev}
\end{figure}

Because most of the orbit uncertainty is in the in-track direction (i.e., 
the timing along the orbit is most uncertain), we compare just the 
marginal posterior standard deviation of the mean longitudes in 
\autoref{fig:thresher_in_track_stddev}. The horizontal dashed line in 
\autoref{fig:thresher_in_track_stddev} shows the standard deviation on the 
mean longitude inferred from the simultaneous ground- and space-based 
observations. The black points show the decrease in the mean longitude 
standard deviation with increasing numbers of consecutive 30-second exposures 
from the single ground-based telescope. 
As in \autoref{fig:thresher_marg_dens_comparison}, we see that 10 exposures 
from the ground yields comparable constraints to the 3 telescope scenario, but
that fewer consecutive exposures leaves significantly larger uncertainties 
on the mean longitude. That is, there is a relatively slow rate of convergence
between the two observing scenarios.

\section{Conclusions}
\label{sec:conclusions}

Using a simulation study, we demonstrated a new algorithm to combine images 
of a satellite from multiple telescopes or exposures for orbit determination 
and refinement. 

Our algorithm uses a probabilistic forward model for satellite 
images in sidereal tracking optical observations. The output of a probabilistic 
fit to an image is samples of streak image model parameters from an interim 
posterior distribution. These samples constitute a form of statistical image 
compression. Because we propagate all uncertainties in the streak fitting 
process we can select images of arbitrary signal-to-noise ratio. 

Given interim posterior samples of streak image parameters from an image, we 
perform a preliminary orbit determination (POD) using a `statistical ranging'
algorithm from \cite{virtanen2001,schneider12} to transform image positions to 
probabilistic samples of osculating Keplerian elements. This is a flexible 
approach to image processing that allows trivial parallelization of computations
across distinct images. 

Given the PODs from all images of a satellite (over all times and telescopes), 
we can combine the statistical samples of orbit parameters with a new 
importance sampling algorithm. The importance sampling requires new evaluations 
of the likelihood functions of each exposure, but can again be massively 
parallelized unlike Markov Chain algorithms~\cite{schneider12}.

In our simulation study we found distinct advantages in orbit determination 
in the combination of ground- and space-based observing deriving from the 
geometry of separate observatories. To perform similar orbit inferences from a 
single ground-based telescope requires a higher cadence of tracking 
observations, which we expect to become impractical as more sources and larger 
areas of the sky are monitored.

\section*{Acknowledgments}
This work was performed under the auspices of the U.S. Department of Energy for 
Lawrence Livermore National Laboratory under Contract DE-AC52-07NA27344. 
Funding for this work was provided by LLNL Laboratory Directed
Research and Development grant 16-ERD-013.

\bibliographystyle{plain}
\bibliography{orbitdetermination}
\end{document}